\documentclass[twocolumn,showpacs,prb,superscriptaddress]{revtex4}

\usepackage{graphicx}
\usepackage{color}
\usepackage{tabularx}
\usepackage{epsfig}
\usepackage{amsmath}
\usepackage{amssymb}
\usepackage{graphicx}
\usepackage{dcolumn}
\usepackage{bm}
\usepackage{wasysym}

\definecolor{grn}{rgb}{0,0,0.54}

\newcommand{\ket}[1]{\ensuremath{| #1 \rangle}}

\begin{document}


\title{Quantum Phase Transition in a Heisenberg Antiferromagnet on a Square Lattice with Strong
Plaquette Interactions}

\author{A. Fabricio Albuquerque}
\affiliation{School of Physics, The University of New South Wales, Sydney, NSW 2052, Australia}

\author{Matthias Troyer}
\affiliation{Theoretische Physik, ETH Zurich, 8093 Zurich, Switzerland} 

\author{Jaan Oitmaa}
\affiliation{School of Physics, The University of New South Wales, Sydney, NSW 2052, Australia}


\date{\today}
\pacs{02.70.-c,02.70.Ss,75.10.Jm,75.40.Mg}

\begin{abstract}
We present numerical results for an $S=1/2$ Heisenberg antiferromagnet on a inhomogeneous
square lattice with tunable interaction between spins belonging to different plaquettes. Employing
Quantum Monte Carlo, we significantly improve on previous results for the the critical point separating
singlet-disordered and N\'{e}el-ordered phases, and obtain an estimate for the critical exponent
$\nu$ consistent with the three-dimensional classical Heisenberg universality class. Additionally, we show
that a fairly accurate result for the critical point can be obtained from a Contractor Renormalization (CORE)
expansion by applying a surprisingly simple analysis to the effective Hamiltonian.
\end{abstract}

\maketitle



\section{Introduction}
\label{sec:intro}

The square lattice quantum Heisenberg antiferromagnet (SLQHA) with spin $S=1/2$ is
one of the paradigmatic models in condensed matter physics and has been extensively
investigated in the last two decades, mainly in connection with the parent compounds of
cuprate superconductors. \cite{manousakis:91} Good agreement with experimental data
is obtained from the analysis of an effective continuous field theory, rigorously justified in the
limit of large spin $S$, given by the $(2+1)$-dimensional nonlinear $\sigma$ model
(NL$\sigma$M).\cite{chakravarty:89} The coupling $g$ in the NL$\sigma$M
controls the transition between N\'{e}el-ordered and quantum disordered phases and it
can be shown that the SLQHA maps to the renormalized classical regime of the
NL$\sigma$M ($g<g_{\rm c}$), which therefore has a long-range ordered ground-state.
\cite{manousakis:91,chakravarty:89}

The proximity to a quantum critical point may, however, lead to a crossover to a finite 
temperature quantum critical regime where thermodynamic quantities are affected by
strong quantum fluctuations and display universal behavior.\cite{chakravarty:89,chubukov:94}
Nevertheless, numerical simulations\cite{kim:98} indicate that this crossover may be too
narrow to be detectable, possibly due to the fact that the SLQHA is deep inside the renormalized
classical regime. This difficulty has motivated the investigation of antiferromagnets defined on
{\em decorated} square lattices where the proximity to a quantum critical point separating
singlet-disordered and N\'{e}el-ordered phases is controlled by adjusting couplings in the
Hamiltonian.\cite{sandvik:94,troyer:96,kotov:98,shevchenko:00,matsumoto:01,wang:06}

\begin{figure}
  \begin{center}
    \includegraphics*[width=0.22\textwidth]{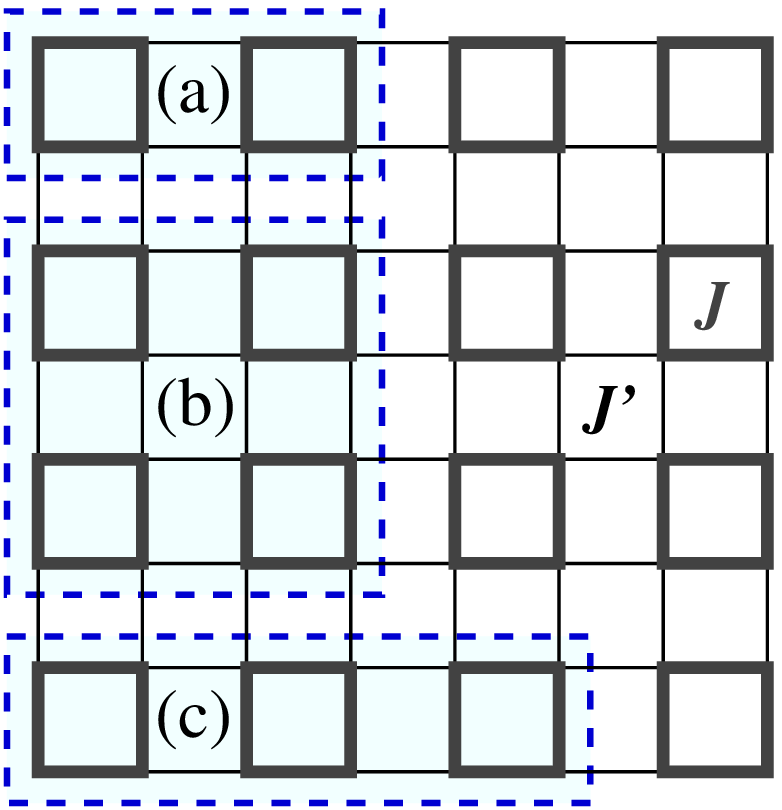}
  \end{center}
  \caption{(Color online) The ``plaquettized" square lattice considered in this paper: nearest-neighbor
  spins lying on the same (neighboring) plaquette(s) interact via superexchange $J$ ($J'$), represented
  by thick (thin) continuous lines [see Eq.~(\ref{eq:hamiltonian})]. Dashed lines highlight the clusters
  employed in obtaining the range-$1$ (a), -$2^{1/2}$ (b) and -$2$ (c) CORE results.}
  \label{fig:Lattice}
\end{figure}

We consider an $S=1/2$ Heisenberg Hamiltonian defined on the ``plaquettized" square
lattice depicted in Fig.~\ref{fig:Lattice}:
\begin{equation}
    {\mathcal H} = 	J \sum_{\left\langle i,j \right\rangle} \vec{S}_{i} \cdot \vec{S}_{j}
    				+ J' \sum_{\left\langle i,j \right\rangle'} \vec{S}_{i} \cdot \vec{S}_{j}~.
  \label{eq:hamiltonian}
\end{equation}
$J$ ($\left\langle i,j \right\rangle$, bold lines in Fig.~\ref{fig:Lattice}) and $J'$
($\left\langle i,j \right\rangle'$, thin lines in Fig.~\ref{fig:Lattice}) are, respectively, intra-
and inter-plaquette nearest-neighbor antiferromagnetic interactions. Since the model is
self-dual under the transformation $J \leftrightarrow J'$, we apply the restriction $J' \leq J$ without
loosing generality and set $J=1$. Similarly to what happens with the aforementioned models
with tunable interactions,\cite{sandvik:94,troyer:96,kotov:98,shevchenko:00,matsumoto:01,wang:06}
the ratio $J'/J$ (equivalent to $g^{-1}$ in the NL$\sigma$M) controls the magnitude of quantum
fluctuations and a quantum phase transition at $J'_{\rm C}/J$, belonging to the universality class
of the 3D classical Heisenberg model,\cite{chakravarty:89} separates a disordered singlet phase
at low $J'/J$ from the renormalized classical state at $J'/J=1$, where the original SLQHA is recovered.
Previous results for this ``plaquettized" Heisenberg antiferromagnet were
obtained analytically,\cite{koga:99a} from series expansions,\cite{koga:99a,singh:99} exact
diagonalization of small clusters \cite{laeuchli:02,voigt:02} and by diagonalizing the effective model
obtained from a Contractor Renormalization (CORE) expansion.\cite{capponi:04,capponi:06} The
currently best estimate for the critical point [$J'_{\rm C}/J=0.555(10)$] was obtained from an Ising
series expansion.\cite{singh:99}

We investigate the model defined by Eq.~(\ref{eq:hamiltonian}) by means of Quantum Monte Carlo
(QMC) simulations and CORE with a twofold purpose. Firstly, we would like to
improve on previous estimates\cite{koga:99a,singh:99,laeuchli:02,voigt:02,capponi:04,capponi:06}
for the quantum critical point. This might pave the way to future investigations of, for instance,
the effects of impurities in a quantum critical antiferromagnet.\cite{sachdev:99b,hoeglund:07a,hoeglund:07b}
Secondly, we are interested in testing the quality of the results obtained from a CORE expansion
including longer-ranged effective interactions than in Refs.~\onlinecite{capponi:04,capponi:06} and
by applying a simpler analysis to the effective Hamiltonian.


\section{Numerical Results}
\label{sec:numerical}

\begin{figure}
  \begin{center}
    \includegraphics*[width=0.3\textwidth,angle=270]{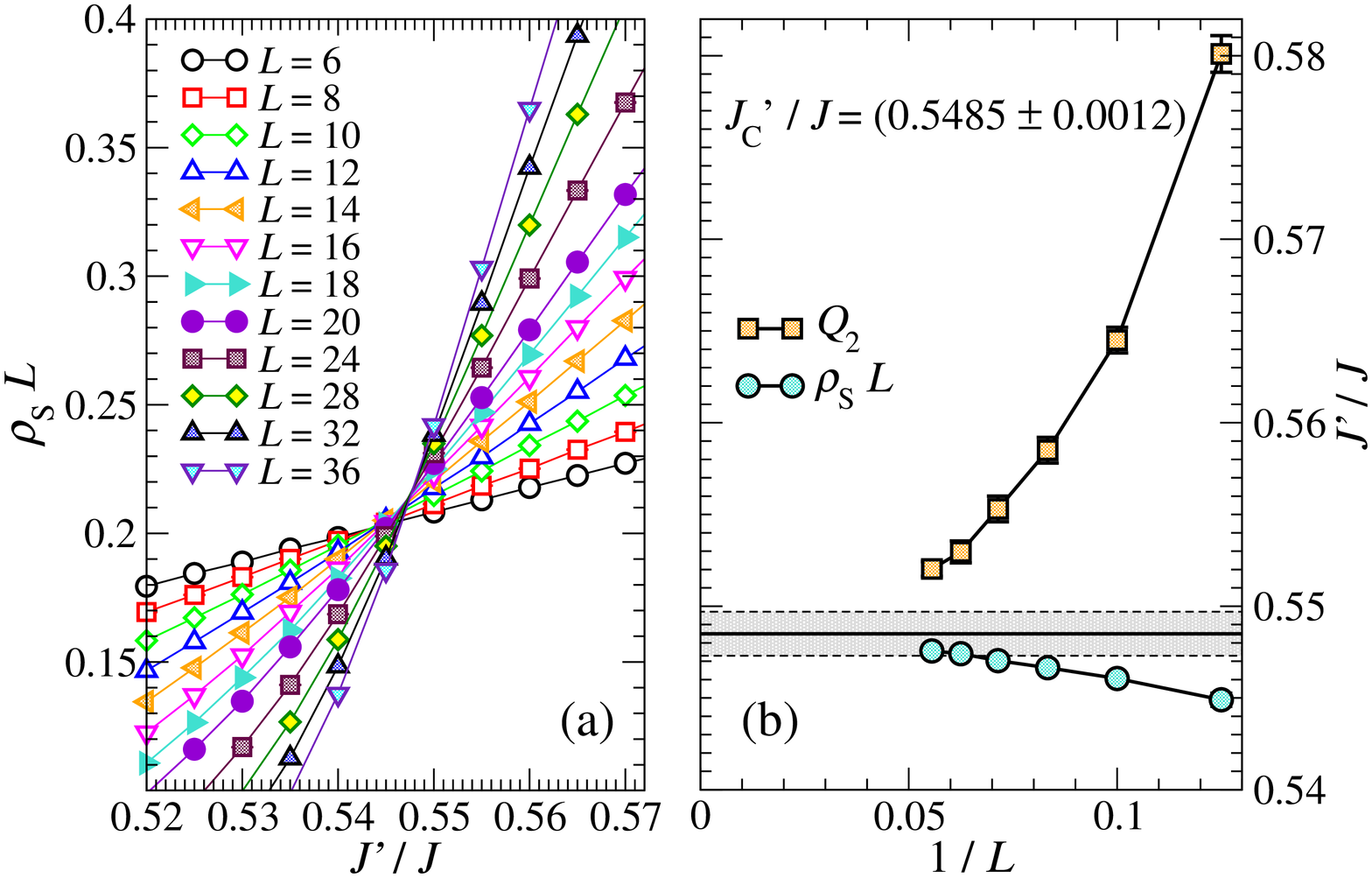}

    \vspace{0.2cm}
 
    \includegraphics*[width=0.18\textwidth,angle=270]{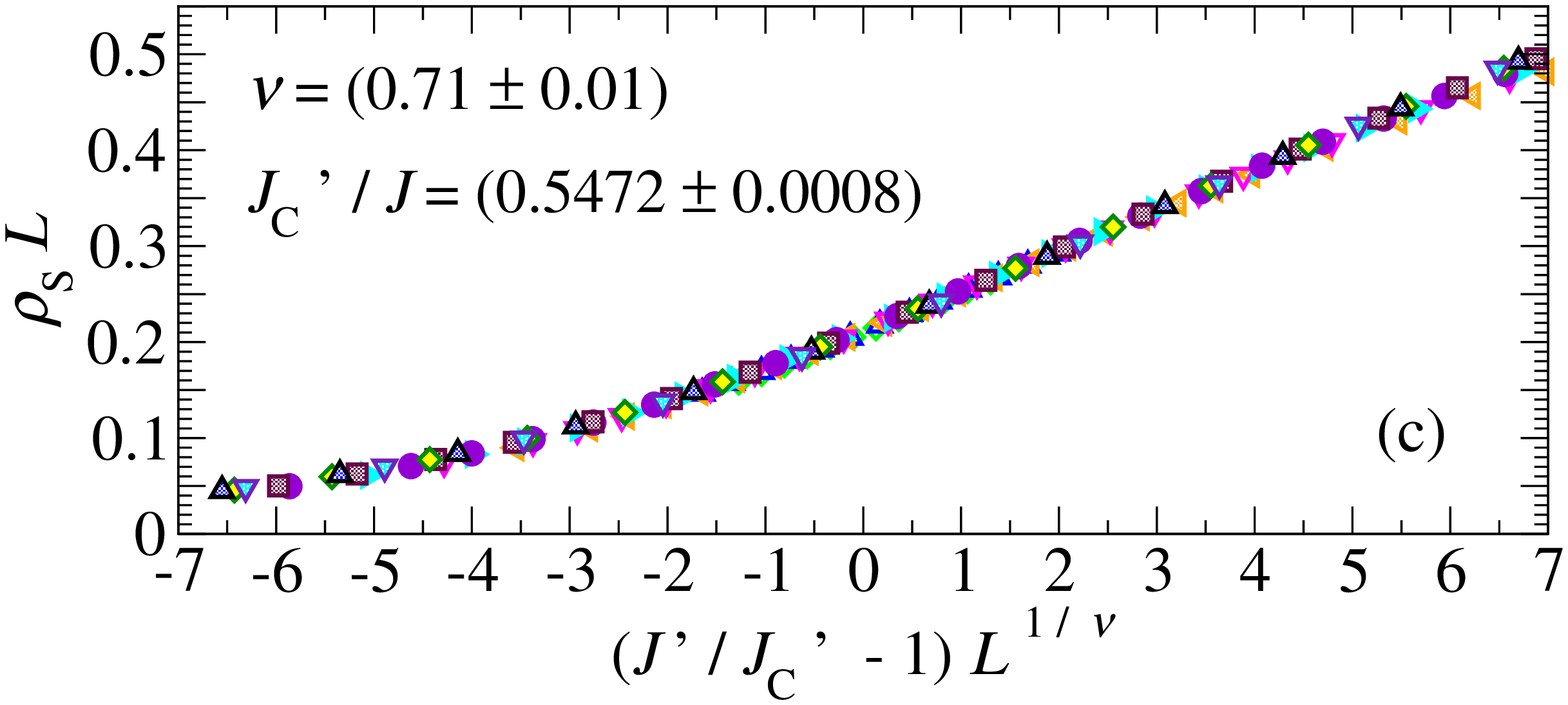}
  \end{center}
  \caption{(Color online) (a) SSE-QMC data for spin stiffness $\rho_S$ multiplied by the system's
  size $L$ as a function of the inter-plaquette coupling $J'/J$, for temperatures $T \leq 1/2L$.
  Error bars are much smaller than the symbols' size. (b) Convergence of the intersection points
  for curves $L$ and $2L$ for $\rho_{S}L$ [data shown in (a)] and second-order Binder cumulant $Q_2$
  (not shown). Extrapolation (see main text) gives us the estimate $J'_{\rm C}/J = 0.5485(12)$
  for the critical point. (c) Data collapse for $\rho_SL$ is attained for $J'_{\rm C}/J = 0.5472(8)$
  and $\nu = 0.71(1)$, using the scaling Ansatz of Eq.~(\ref{eq:scaling}).}
  \label{fig:QMC}
\end{figure}
\subsection{Quantum Monte Carlo}
\label{sec:QMC}

We have performed QMC simulations for the model described by
Eq.~(\ref{eq:hamiltonian}) by employing the ALPS libraries\cite{albuquerque:07} implementation
of the {\em directed loops} algorithm\cite{syljuasen:02,alet:05:a} for the Stochastic Series Expansion (SSE)
representation.\cite{sandvik:99}
Lattices with $L \times L$ sites ($L/2 \times L/2$ plaquettes) with $L$ up to $36$ have been considered,
with periodic boundary conditions along both directions. Temperatures are set so to ensure
that ground-state properties are accessed: for each system's size, simulations were performed for
inverse temperatures $\beta_n = 2^n$ so that the obtained averages agreed (within error bars)
for $\beta_n$ and $\beta_{n-1}$.\cite{beta} We have calculated the spin stiffness $\rho_{\rm S}$ and
the second-order Binder cumulant for the staggered magnetization, $Q_{2}$. The spin stiffness is
obtained in terms of the winding numbers $w_{\rm x}$ and $w_{\rm y}$,
\begin{equation}
  \rho_{\rm S}=\frac{1}{2\beta L^2} \langle {w_{\rm x}}^2+{w_{\rm y}}^2\rangle~,
  \label{eq:rho}
\end{equation}
and is expected to scale close to the critical point like $\rho_{\rm S} \sim L^{2-d-z}$, where $d = 2$ is the
dimensionality and the dynamic critical exponent is expected to be $z=1$.\cite{fisher:89} Therefore the
quantity $\rho_{\rm S}L$ should assume a size-independent value at the critical point, something confirmed
by our results shown in Fig.~\ref{fig:QMC}(a). The second-order Binder cumulant for the staggered
magnetization ($m_{\rm s}^{z}$) is defined as
\begin{equation}
  Q_{2}= \frac{\langle (m_{\rm s}^{z})^{4} \rangle}{\langle (m_{\rm s}^{z})^{2} \rangle^{2}}~.
  \label{eq:binder}
\end{equation}
$Q_{2}$ also displays universal behavior in the critical regime and curves for different lattice sizes cross
close to the critical point (not shown).

In order to locate the quantum critical point, $J'_{\rm C}/J$, we first analyze the scaling behavior
of the intersection points between curves for
$\rho_{\rm S}L$ and $Q_2$ obtained for lattice sizes ($L$, $2L$). Crossing points are
determined by performing linear and quadratic fits to different data subsets and deviations between
different estimates are used in setting (generous) error bars. The so obtained results are plotted
in Fig.~\ref{fig:QMC}(b) as a function of $1/L$. We remark, similarly to what was found in
Ref.~\onlinecite{wang:06} for dimerized magnets, that fastest convergence to the thermodynamic limit
is attained for $\rho_{\rm S}L$ and therefore focus on the results for this quantity in what follows.\cite{binder}
Since the curvature for the crossing points decreases with increasing $L$ [Fig.~\ref{fig:QMC}(b)], an upper bound for
$J'_{\rm C}/J$ is simply obtained by applying a linear extrapolation to the three largest ($L$, $2L$)
crossing points; a lower-bound is directly given by the crossing point for the largest pair ($L$, $2L$).
In this way, we arrive at the result $J'_{\rm C}/J = 0.5485(12)$.\cite{binder}

In trying to achieve higher accuracy, and additionally estimate the critical exponent associated to the
correlation length, $\nu$, we employ the scaling Ansatz
\begin{equation}
  \rho_{\rm S}(t,L) = L^{-1}f_{\rho_{\rm S}} (tL^{1/\nu})~,
  \label{eq:scaling}
\end{equation}
with reduced coupling $t = (J' - J'_{\rm C}) / J'_{\rm C}$. By plotting $\rho_{\rm S}L$
versus $tL^{1/\nu}$, and adjusting the values of $J'_{\rm C}/J$ and $\nu$, we achieve data
collapse for $J'_{\rm C}/J = 0.5472(8)$ and $\nu = 0.71(1)$, as shown in Fig.~\ref{fig:QMC}(c).\cite{binder}

Our results for $J'_{\rm C}/J$ are consistent with earlier estimates,\cite{koga:99a,singh:99,laeuchli:02,
voigt:02,capponi:04,capponi:06} but improve on the previously best result [$J'_{\rm C}/J=0.555(10)$,
Ref.~\onlinecite{singh:99}] by one order of magnitude. Furthermore, our estimate for the critical exponent $\nu$
is compatible with the most accurate result for the 3D classical Heisenberg model [$\nu=0.7112(5)$,
Ref.~\onlinecite{campostrini:02}], as expected from the mapping onto a NL$\sigma$M.\cite{chakravarty:89}
Further improvements in the values for $J'_{\rm C}/J$ and $\nu$ may be achieved by
simulating larger systems and/or applying a more sophisticated data analysis, taking into account
subleading finite-size corrections.\cite{beach:05,wang:06}


\begin{figure}
  \begin{center}
    \includegraphics*[width=0.33\textwidth,angle=270]{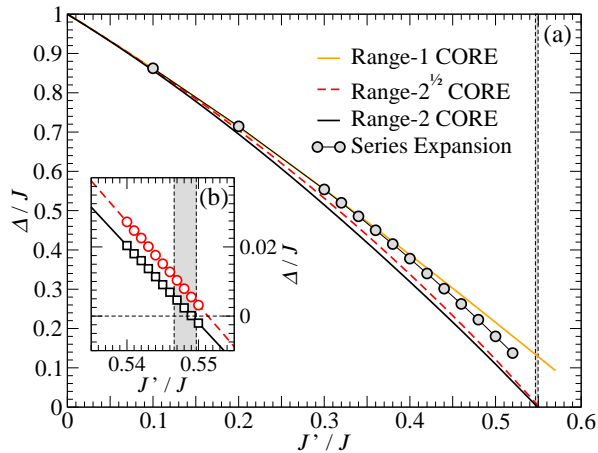}
  \end{center}
  \caption{(Color online) (a) Spin-gap in the disordered phase of the spin model on the modulated square
  lattice described by Hamiltonian Eq.~(\ref{eq:hamiltonian}). Results have been obtained from a
  plaquette series expansion\cite{series} (circles, Ref.~\onlinecite{singh:99}) and various range CORE analysis (see main text).
  The vertical dashed lines indicate values of $J'/J$ consistent with the critical point obtained from QMC simulations
  ($J'_{C}/J \in [0.5466,0.5497]$). The zoom in the inset (b) shows that values for the critical point consistent with the
  ones obtained from QMC are obtained from range-2 (squares) CORE results (range-$2^{1/2}$ results are represented
  by red circles).}
  \label{fig:CORE}
\end{figure}

\subsection{Contractor Renormalization}
\label{sec:CORE}

The Contractor Renormalization (CORE) method\cite{morningstar:94,morningstar:96} is a tool
in deriving low-energy effective Hamiltonians for lattice models and was previously applied to
the spin Hamiltonian on the modulated square lattice [Eq.~(\ref{eq:hamiltonian})] by Capponi
{\em et al.}.\cite{capponi:04,capponi:06} We extend their results by deriving a longer-ranged
CORE expansion. We also notice that our CORE expansion is essentially a strong coupling
(on-site repulsion $U \rightarrow \infty$) version of the one derived in Ref.~\onlinecite{altman:02},
where an effective Hamiltonian for the Hubbard model on the square lattice was obtained, but we
have the advantages of a more natural motivation for choosing the plaquettes as elementary blocks
and of the absence of charge degrees of freedom.

We start by noticing that the original spin model described by Eq.~(\ref{eq:hamiltonian}) is
particularly amenable to a CORE analysis, for the strongly coupled plaquettes are a natural
choice as the elementary blocks (see Fig.~\ref{fig:Lattice}; for details on the CORE procedure the
reader is referred to Refs.~\onlinecite{capponi:06} and \onlinecite{altman:02}). The retained low-lying
block states are the plaquette's singlet ground-state $\ket{s}$ and the triplet states $\ket{t^{\alpha}}$, with
$\alpha = -1, 0, +1$ denoting the total $S^z$ component. This choice for the restricted local basis
is justified by the fact that $\ket{s}$ and $\ket{t^{\alpha}}$ are the lowest block's eigenstates and also,
as shown by Capponi {\em et al.},\cite{capponi:04,capponi:06} by their large weight in the
density matrix of a plaquette embedded in a larger cluster. Effective couplings between these retained
block states are obtained by subsequently diagonalizing a cluster comprised of connected plaquettes:
a matching number of cluster's low-lying eigenstates are projected upon the basis formed by the tensor
products of local singlets and triplets, an effective Hamiltonian being obtained by imposing the constraint
that the cluster's low-energy spectrum is exactly reproduced and by subtracting interactions previously
obtained from clusters involving lesser plaquettes. We employ the clusters highlighted in
Fig.~\ref{fig:Lattice} and label the results according to the longest range effective couplings obtained
at each step of the CORE expansion: range-$1$ interactions are obtained from the cluster comprised by
two connected plaquettes depicted in Fig.~\ref{fig:Lattice}(a), range-$2^{1/2}$ from a four-plaquette cluster
[Fig.~\ref{fig:Lattice}(b)] and range-$2$ from the cluster displaying three aligned plaquettes shown in
Fig.~\ref{fig:Lattice}(c).

The effective model resulting from the above procedure is expected to evidence dominant microscopic
mechanisms at play in the original model and physically sound results are ideally obtained by
means of a simplified subsequent analysis. We stress that our approach differs from the previous one
\cite{capponi:04,capponi:06} in a crucial way: while Capponi {\em et al.}\cite{capponi:04,capponi:06}
restricted their CORE expansion to the shortest-range and studied the resulting range-1 effective
Hamiltonian by means of exact diagonalizations, we expect that good results are obtainable in a simpler
way from an extended CORE expansion including longer-ranged interactions. Accordingly, we locate the
quantum critical point by determining the value of $J'/J$ where the gap for triplet
excitations ($\Delta$) vanishes and estimate $\Delta$ from the effective CORE Hamiltonian simply as
\begin{equation}
    \Delta= 	\mu_{\rm t} - 4 (t_1 + t_2 + t_3)~.
  \label{eq:Gap_CORE}
\end{equation}
$\mu_{\rm t}$ is the chemical potential for triplet excitations above the singlet ground-state for low $J'/J$
and the second term accounts for the triplons' kinetic energy: $t_1$ is the nearest-neighbor (NN), $t_2$ the
next-NN and $t_3$ the third-NN hopping amplitudes.\cite{CORE-range} In other words,
Eq.~(\ref{eq:Gap_CORE}) is the energy of an {\em isolated} triplon in a {\em singlet sea}.

Results for $\Delta(J'/J)$ obtained from range-1, -$2^{1/2}$ and -2 CORE expansions are shown in
Fig.~\ref{fig:CORE}, compared with the ones obtained from the plaquette series expansion derived in
Ref.~\onlinecite{singh:99}.\cite{series} As expected, all results mutually agree in the limit of small $J'/J$,
where they are essentially exact; for larger $J'/J$, range-$2^{1/2}$ and -2 CORE
underestimate $\Delta$. However, the values of $J'/J$ where range-$2^{1/2}$ and -2 results for
$\Delta$ vanish are in surprisingly good agreement with the QMC results for $J'_{\rm C}/J$
presented in Sec.~\ref{sec:QMC}: $J'/J \approx 0.5513$ for range-$2^{1/2}$ and
$J'/J \approx 0.5491$ for range-2. These values are seemingly converging very fast to a result
consistent with the QMC estimates and this suggests that the procedure employed here
might lead to more precise estimates for the critical point than the one employed in
Refs.~\onlinecite{capponi:04} and \onlinecite{capponi:06}, where $J'_{\rm C}/J=0.55(5)$ was
obtained. We conjecture that this unexpected accuracy is related to
level-crossings observed close to the point where $\Delta$ vanishes. However, obviously
we cannot discard the possibility that this remarkable agreement is coincidental and remark
that poorer results are obtained for intermediate values of $J'/J$, with the consequence that
$\Delta$ from CORE does not follow a power law (as seen from a logarithmic plot, not shown).
However, it is desirable to further test the procedure employed here by considering
similar spin models.\cite{sandvik:94,troyer:96,kotov:98,shevchenko:00,matsumoto:01,wang:06}


\section{Conclusions}

Summarizing, we have investigated an $S=1/2$ Heisenberg antiferromagnet on a ``plaquettized"
square lattice (Fig.~\ref{fig:Lattice}) by means of QMC and CORE. Our results for the quantum
critical point separating the gapped-singlet and N\'{e}el-ordered phases, $J'_{\rm C}/J = 0.5485(12)$ and
$J'_{\rm C}/J = 0.5472(8)$, obtained from QMC simulations, substantially improve on previous estimates,
\cite{koga:99a,singh:99,laeuchli:02,voigt:02,capponi:04,capponi:06} and the obtained
critical exponent $\nu = 0.71(1)$ is consistent with the tridimensional classical Heisenberg model universality
class,\cite{campostrini:02} as expected from the mapping to a NL$\sigma$M.\cite{chakravarty:89}

We also highlight the surprisingly good result for the critical point extracted from a simple analysis
of range-$2^{1/2}$ and -2 effective CORE Hamiltonians. However, it is not presently possible to exclude
the possibility that the good agreement with QMC is coincidental and it would be interesting to further
test the procedure employed here. The fact that CORE is immune to the infamous {\em sign problem}
open interesting research possibilities, and fermionic systems on a geometry similar to the one considered
here\cite{trebst:06} may be investigated.

{\bf Note:} While preparing this manuscript, and after finishing our simulations, we became aware of work
by Wenzel {\em et al.}\cite{wenzel:08a,wenzel:08b} in which dimerized and quadrumerized two-dimensional
antiferromagnets are investigated by QMC. By employing a more sophisticated data analysis and simulating
larger lattices, they obtain more precise estimates for the critical
point for the model considered in the present work.


\begin{acknowledgments}
We thank C.~J.~Hamer and O.~P.~Sushkov for fruitful discussions.
  QMC simulations were performed on the clusters Hreidar and Gonzales at ETH-Zurich.
  This work has been supported by the Australian Research Council.
\end{acknowledgments}

\bibliographystyle{apsrev}

\begin{thebibliography}{36}
\bibitem[{bet()}]{beta}
\bibinfo{note}{Convergence was tested for values of $J'/J$ close to the
  critical point, as estimated from preliminary QMC runs.}
\bibitem[{bin()}]{binder}
\bibinfo{note}{Results obtained from the analysis of the Binder cumulant,
  $J'_{\rm C}/J = 0.552(4)$ and $\nu = 0.75(5)$, are consistent with the ones
  obtained from $\rho_{\rm S}L$ but are less precise, seemingly due to stronger
  subleading finite-size corrections.}
\bibitem[{COR()}]{CORE-range}
\bibinfo{note}{The actual expression for $\Delta$, Eq.~(\ref{eq:Gap_CORE}),
  changes according to the range of the CORE expansion: range-1 $\Delta$ is
  determined by $\mu_{\rm t}$ and $t_1$ only, for range-$2^{1/2}$, $t_2$ is
  also included and for range-2 the full expression is recovered. At each step,
  corrections for $\mu_{\rm t}$ and $t_1$ are taken into account.}
\end{thebibliography}

\end{document}